\documentclass[aps,prl,twocolumn,superscriptaddress,longbibliography]{revtex4-1}%
\usepackage{amsmath,amssymb}
\usepackage{graphicx}
\usepackage{epstopdf}
\usepackage{bm}
\usepackage{color}
\usepackage{amsmath}
\usepackage{amsfonts}
\usepackage{amssymb}%
\providecommand{\U}[1]{\protect\rule{.1in}{.1in}}

\begin{document}
\title{Frustration-driven C$_{4}$ symmetric orders in a hetero-structured iron-based superconductor}
\author{Jong Mok Ok}
\affiliation{Department of Physics, Pohang University of Science and Technology, Pohang
790-784, Korea}
\affiliation{Center for Artificial Low Dimensional Electronic Systems, Institute for Basic
Science, Pohang 790-784, Korea}
\author{S.-H. Baek}
\email{[Corresponding author:]sbaek.fu@gmail.com}
\affiliation{IFW Dresden, Helmholtzstr. 20, 01069 Dresden, Germany}
\author{C. Hoch}
\affiliation{Max-Planck-Institut f\"{u}r Festk\"{o}rperforschung, Heisenbergstra$\beta$e 1,
D-70569 Stuttgart, Germany}
\author{R. K. Kremer}
\affiliation{Max-Planck-Institut f\"{u}r Festk\"{o}rperforschung, Heisenbergstra$\beta$e 1,
D-70569 Stuttgart, Germany}
\author{S. Y. Park}
\affiliation{Max Planck POSTECH Center for Complex Phase Materials, Pohang University of
Science and Technology, Pohang 790-784, Korea}
\author{Sungdae Ji}
\affiliation{Department of Physics, Pohang University of Science and Technology, Pohang
790-784, Korea}
\affiliation{Max Planck POSTECH Center for Complex Phase Materials, Pohang University of
Science and Technology, Pohang 790-784, Korea}
\author{B. B\"uchner}
\affiliation{IFW Dresden, Helmholtzstr. 20, 01069 Dresden, Germany}
\author{J.-H. Park}
\affiliation{Department of Physics, Pohang University of Science and Technology, Pohang
790-784, Korea}
\affiliation{Max Planck POSTECH Center for Complex Phase Materials, Pohang University of
Science and Technology, Pohang 790-784, Korea}
\affiliation{Division of Advanced Materials Science, Pohang University of Science and
Technology, Pohang 790-784, Korea}
\author{S. I. Hyun}
\affiliation{Department of Chemistry, Pohang University of Science and Technology, Pohang
790-784, Korea}
\author{J. H. Shim}
\affiliation{Department of Chemistry, Pohang University of Science and Technology, Pohang
790-784, Korea}
\author{Yunkyu Bang}
\affiliation{Department of Physics, Chonnam National University, Korea}
\author{E. G. Moon}
\affiliation{Department of Physics, Korea Advanced Institute of Science and Technology,
Daejeon 305-701, Korea}
\author{I. I. Mazin}
\affiliation{Naval Research Laboratory, code 6390, 4555 Overlook Avenue S.W., Washington,
DC 20375, USA}
\author{Jun Sung Kim}
\email{[Corresponding author:]js.kim@postech.ac.kr}
\affiliation{Department of Physics, Pohang University of Science and Technology, Pohang
790-784, Korea}
\affiliation{Center for Artificial Low Dimensional Electronic Systems, Institute for Basic
Science, Pohang 790-784, Korea}
\date{\today}

\begin{abstract}
A subtle balance between competing interactions in strongly correlated systems can be easily tipped by additional interfacial interactions in a heterostructure. This often induces exotic phases with unprecedented properties, as recently exemplified by high-$T_c$ superconductivity in FeSe monolayer on the nonmagnetic SrTiO$_{3}$. When the proximity-coupled layer is magnetically active, even richer phase diagrams are expected in iron-based superconductors (FeSCs), which however has not been explored due to the lack of a proper material system. One promising candidate is Sr$_{2}$VO$_{3}$FeAs, a
naturally-assembled heterostructure of a FeSC and a Mott-insulating vanadium oxide. Here, using high-quality single crystals and high-accuracy $^{75}$As\ and $^{51}$V\ nuclear magnetic resonance (NMR) measurements, we show that a novel electronic phase is emerging in the FeAs layer below $T_{0}\sim155$ K without either static magnetism or a crystal symmetry change, which has never been observed in other FeSCs. We find that frustration of the otherwise dominant Fe stripe and V Neel fluctuations via interfacial coupling induces a charge/orbital order with $C_{4}$-symmetry in the FeAs layers, while suppressing the Neel antiferromagnetism in the SrVO$_{3}$ layers. These findings demonstrate that the magnetic proximity coupling is effective to stabilize a hidden order in FeSCs and, more generally, in strongly correlated heterostructures.
\end{abstract}
\maketitle

In strongly correlated electron materials, including cuprates, transition metal oxides (TMOs), and iron-based superconductors (FeSCs), competing interactions of spin, charge and orbital degrees of freedom lead to complex and rich phase diagrams, extremely sensitive to external perturbations. Especially impressive is modification of the phase diagram via introducing interfacial interactions, as intensively studied for the heterostructures of high-$T_c$ cuprates\cite{gozar08,wu13,chakhaklian06,sataphthy13,driza13} or transition metal oxides\cite{hwang12,chakhaklian14}, showing the enhanced $T_c$ or new emergent phases that cannot be stabilized in their constituent layer alone. The similar effect has also been found in FeSCs, for example, in FeSe monolayers on top of nonmagnetic SrTiO$_3$~\cite{tan13,ge15,lee14} showing drastically enhanced $T_c$, arguably higher than 100 K (Ref. \cite{ge15}). Although the underlying mechanism is yet to be confirmed, the interfacial coupling is considered to be critical and may further enhance $T_c$ in the superlattice~\cite{coh16}. Of particular interest is when the proximity coupled layer is strongly correlated and magnetically active. As found in heterostructures of high-$T_c$ cuprates and magnetic TMOs~\cite{chakhaklian06,sataphthy13,driza13}, additional interfacial spin interaction may also induce novel ground states of FeSCs in proximity of a Mott insulator, which however has not been explored so far.

Sr$_{2}$VO$_{3}$FeAs is a unique member of the family of FeSCs, a very rare \emph{naturally-assembled} superlattice of [SrFeAs]$^{+1}$ and [SrVO$_{3}$]$^{-1}$ layers\cite{zhu09a}, as shown in Fig. 1a. Initially Sr$_{2}$VO$_{3}$FeAs was thought to have, because of the V bands, an unusual Fermi surface topology, incompatible with $s^{\pm}$ superconductivity~scenario driven by spin-fluctuation~\cite{lee10b}. However, it was soon realized the V $3d^{2}$ electrons in the SrVO$_{3}$ layer are strongly correlated and form a Mott-insulating state~\cite{qian11a,nakamura10,kim15a}, while the partially-filled Fe $3d^{6}$ state in the FeAs layer has the considerable itinerancy and superconducts at $T_{c}\sim35$ K~\cite{zhu09a,lee10b,mazin10a}. These contrasting ground states in Sr$_{2}$VO$_{3}$FeAs\ make this system prototypical for strongly correlated heterostructures based on FeSCs and TMOs. Sr$_{2}$VO$_{3}$FeAs has the Fermi surface similar to that in other FeSCs~\cite{qian11a,kim15a}, and thus is expected to show either the stripe antiferromagnetic (AFM) order with the wave vector $\mathbf{Q}$=$(\pi,0)$, or the corresponding nematic phase, or enhanced spin fluctuations at low temperature with the same~\cite{johnston10}. There is in fact a second-order transition observed at $T_{0}\sim155$ K with a sizable entropy loss of $\sim$ 0.2$R$ln2 ($R$ is the gas constant)~\cite{sefat11,cao10,tatematsu10}. With no evidence of a static magnetic order or another apparent symmetry breaking, the \emph{hidden} nature of this phase transition, similar to the famous \textquotedblleft hidden-order\textquotedblright\ in underdoped cuprates or a heavy fermion system URu$_{2}$Si$_{2}$, remains elusive and controversial~\cite{sefat11,cao10,tatematsu10,tegel10,hummel13,munevar11,ueshima14}, posing a challenge to our understanding of the physics of FeSCs in proximity of a Mott insulator.

Here we report that an emergent electronic phase is developed below $T_0$ = 155 K in Sr$_2$VO$_3$FeAs, which
is highly distinct in nature from the transitions found in other FeSCs. Using high-accuracy  $^{75}$As\ and $^{51}$V NMR measurements on high-quality single crystals under various field orientations, we unambiguously show that the transition occurs in the FeAs layer, not the SrVO$_3$ layer, without breaking either time reversal symmetry and the underlying tetragonal lattice symmetry. This implies that the typical stripe AFM and $C_2$ nematic phases in the FeAs layers as well as the Neel antiferromagnetism in the SrVO$_3$ layer are significantly suppressed by the interfacial coupling between itinerant iron electrons and localized vanadium spins. We propose that the newly-observed phase is a $C_4$-symmetric charge/orbital order, which has never been observed iron or vanadium-based materials, triggered by frustration of the otherwise dominant Fe stripe and V Neel fluctuations. Our discovery, therefore, offers a new avenue to explore hidden phases with unprecedented properties in the proximity-coupled FeSCs or other strongly correlated electron systems in heterostructures.
\begin{figure}[ptb]
\centering
\includegraphics[width=1\linewidth]{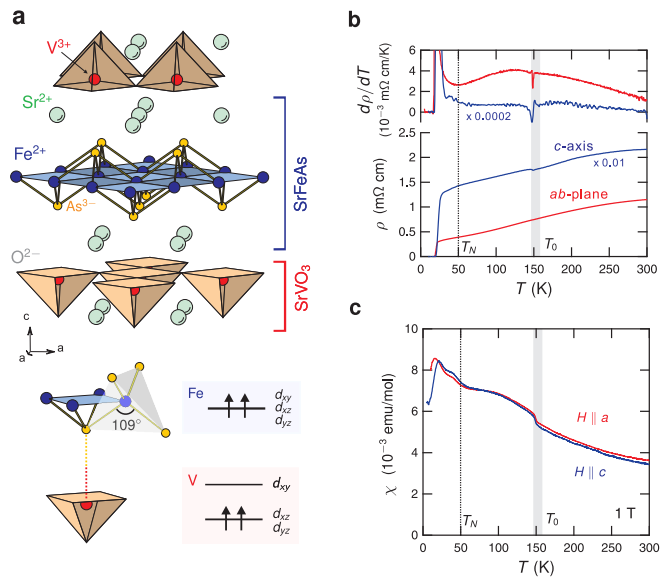}\caption{\textbf{Basic
properties of Sr$_{2}$VO$_{3}$FeAs.} \textbf{\sffamily a}, The crystal
structure of Sr$_{2}$VO$_{3}$FeAs\ as a naturally-assembled heterostructure of
the [SrFeAs]$^{+1}$ and [SrVO$_{3}$]$^{-1}$ layers. V ions form a network of
corner-sharing tetrahedra, while FeAs layers
consist of edge-sharing FeAs$_{4}$ tetrahedra as in other iron-based
superconductors. The local structure of a FeAs$_{4}$ tetrahedron and a
VO$_{3}$ pyramid is highlighted at the bottom. FeAs$_{4}$ tetrahedra provide a
moderate cubic crystal field splitting, much smaller than the band widths,
while the VO$_{3}$ unit is missing one O entirely, and thus develops a strong
Jahn-Teller splitting. The $d_{xy}$ orbital is pushed up, and the two V $d$
electrons occupy the $d_{xz}$ and $d_{yz}$ states, forming an $S=1$
spin. The Fe and the V planes are bridged by the As atoms, as indicated by the
dashed line. \textbf{\sffamily b}, The resistivity $\rho(T)$ in the $ab$ plane and along the $c$ axis shows a weak
anomaly at $T_{0}\sim155$ K, which is more clear in their temperature
derivatives (top panel). The large $c$-axis resistivity, which was scaled down
by a factor of 100, is consistent with the quasi-2D nature.
\textbf{\sffamily c}, The magnetic susceptibility $\chi$($T$), taken at $H=1$
T for $H \perp c$ and $H \parallel c$, shows clear anomalies at $T_{0}$ and
also at $T_{N}\sim45$ K.}
\label{fig:1}
\end{figure}

{\bf Transport and magnetic properties.} Our transport and magnetic measurements on high-quality single crystal of
Sr$_{2}$VO$_{3}$FeAs shown in Figs. 1b and 1c confirm that the transition at
$T_{0}$ is intrinsic. A weak, but discernible, anomaly is observed at
$T_{0}\sim155$ K in the resistivity ($\rho$), even more pronounced in its
temperature derivative $d\rho$/$dT$. The magnetic susceptibility $\chi(T)$
also shows an anomaly at $T_{0}$. Above $T_{0}$, $\chi(T)$ is several times
larger than in typical FeSCs and follows the Curie-Weiss law with a
Curie-Weiss temperature $T_{\text{CW}}\sim-100$ K (see Supplementary section
2). The effective magnetic moment is consistent with $S=1$ expected for the
V$^{3+}$ ions, suggesting that $\chi(T)$ is dominated by localized V spins. At
$T_{0}\sim155$ K, $\chi(T)$ for both $H\parallel ab$ and $H\parallel c$
exhibits a small jump, which corresponds to $\sim$ $10^{-3}$ $\mu_{B}$/f.u.,
three orders of magnitude smaller than typical values of V$^{3+}$ ions
($\sim1.8\mu_{B}$) in vanadium oxides~\cite{nakamura10} and Fe ions
($\sim0.8\mu_{B}$) in FeSCs~\cite{huang08}. Such weak anomalies in $\rho(T)$
and $\chi(T)$, in contrast to a strong one in the specific
heat~\cite{sefat11,cao10,tatematsu10}, question the previous conjectures of a
long-range ordering of either V or Fe spins~\cite{sefat11,cao10,tatematsu10,tegel10,hummel13,munevar11,ueshima14}, and suggest that this weak
ferromagnetic response is only a side effect of the true transition.
However, another anomaly at $T_{N}\sim45$ K in both $\chi_{ab}(T)$ and
$\chi_{c}(T)$ turns out to reflect a long-range ordering of Fe, but still not
V spins, as discussed below. Notably, neither transition is consistent with
the typical stripe AFM or nematic orders for FeSCs.

\begin{figure*}[ptb]
\centering
\includegraphics[width=17cm]{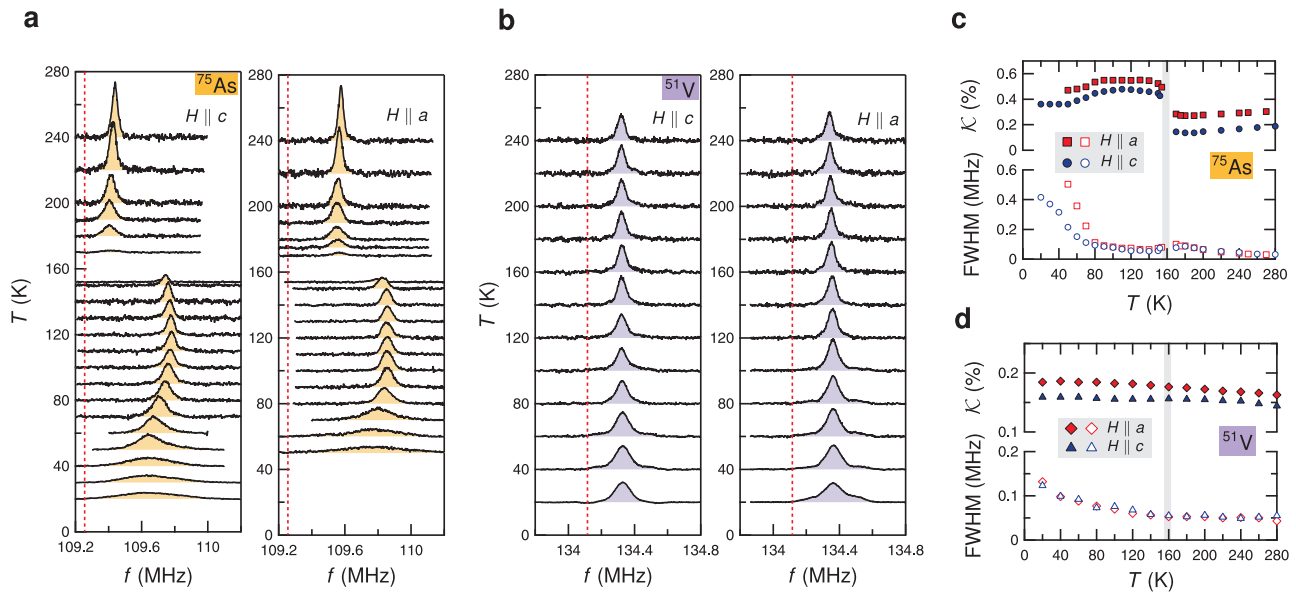}\caption{\textbf{NMR spectra and
their analysis for the Sr$_{2}$VO$_{3}$FeAs\ single crystal.} $^{75}%
$As\ (\textbf{\sffamily a}) and $^{51}$V\ (\textbf{ \sffamily b}) NMR spectra
as a function of temperature, measured in $H=15$ and 12 T, respectively, for the
fields oriented along the $a$ and $c$ axes. The unshifted Larmor frequency
($\nu_{0}\equiv\gamma_{N} H$) is marked by the red vertical lines. Whereas the
$^{51}$V\ spectrum is nearly temperature independent down to 20 K, the $^{75}%
$As\ spectra in both field directions show a sudden shift at $T_{0}\sim155$ K.
\textbf{\sffamily c}, \textbf{\sffamily d}, Temperature dependences of the
$^{75}$As\ and $^{51}$V\ spectra in terms of the Knight shift ($\mathcal{K}$)
and the FWHM, respectively. Below $T_{0}$, a nearly isotropic large jump of
the $^{75}$As\ Knight shift takes place without any magnetic line broadening,
contrasting with the $^{51}$V\ spectra that remain unchanged.}
\label{fig:2}
\end{figure*}

{\bf $^{75}$As\ and $^{51}$V nuclear magnetic resonance spectroscopy.}
To gain further insight into the transition at $T_{0}$ on a microscopic level,
we measured NMR on $^{75}$As\ and $^{51}$V\ nuclei as a function of
temperature for field orientations parallel to $a$ (100), $c$ (001) and the
(110) directions (Fig. 2 and the supplementary Fig. S4). The $^{51}$V\ probes
the V spin order directly and the $^{75}$As\ is a proxy for the Fe sites,
which allows us to probe the two magnetic ions separately.
A dramatic change of the $^{75}$As\ line occurs near $T_{0}\sim155$ K as shown
in Fig. 2a, consistent with the anomalies in $\rho(T)$ and $\chi(T)$. Near 180
K, the $^{75}$As\ signal starts to lose its intensity rapidly and is not
detectable between 150--170 K due to the shortening of the spin-spin
relaxation time $T_{2}$ (Ref. \onlinecite{ueshima14}). Strikingly, the signal
recovers below $\sim150$ K at substantially higher frequencies, in a similar
fashion for both field orientations, which contrasts the typical behaviors of
$^{75}$As\ NMR found in other FeSCs (see Supplementary Fig. S5). This is
better shown in terms of the Knight shift $^{75}\mathcal{K}\equiv(f-\nu
_{0})/\nu_{0}$, where $\nu_{0}\equiv\gamma_{n}H$ with the nuclear gyromagnetic
ratio $\gamma_{n}$ (see Fig. 2\textbf{c}). $^{75}\mathcal{K}$ changes abruptly
at $T_{0}\sim155$ K without any peak splitting or broadening of the full-width
at half-maximum (FWHM) across $T_{0}$. Conversely, the $^{51}$V\ line barely
shifts below $T_{0}$ and down to 20 K (Figs. 2b and 2d), while its FWHM
gradually increases below $T_{0}$. The nearly unchanged $^{51}$V\ NMR line
signals that the V spins remain disordered down to low temperatures. This
contrasting behavior of the $^{75}$As\ and $^{51}$V\ spectra unambiguously
proves that the transition at $T_{0}$ occurs in the FeAs layer, not in the
SrVO$_{3}$ layer, contrary to previous
claims\cite{sefat11,cao10,tatematsu10,tegel10,hummel13,munevar11}.

Having established that the phase transition at $T_{0}$ occurs in the FeAs
layer,
we examined the low-energy Fe spin dynamics, as probed by the $^{75}%
$As\ spin-lattice relaxation rate $T_{1}^{-1}$, which reflects local spin
fluctuations. As shown in Fig. 3, at $T$ $\gtrsim$ 240 K, $(T_{1}T)^{-1}%
$\ exhibits a typical Curie-Weiss-like behavior with an anisotropy
$T_{1,a}^{-1}/T_{1,c}^{-1}$ $\approx$ 1.5. This is expected for a
directionally-disordered state with local stripe AFM correlations with
$\mathbf{Q}=(\pi,0)$ and has been observed in many
FeSCs~\cite{kitagawa08,fu12} (see Supplementary section 6).
With lowering temperature, a critical slowdown of the ($\pi$, 0) spin
fluctuations usually condenses into the $C_{2}$ stripe AFM phase.
For Sr$_{2}$VO$_{3}$FeAs, however, this critical growth is arrested at
$T$$\sim$ 200 K, showing a broad peak of $(T_{1}T)^{-1}$\ with an unusually
large $T_{1,a}^{-1}/T_{1,c}^{-1}$ $\approx$ 6, and then the fluctuations
harden all the way down to $T_{0}$. Across $T_{0}$, $(T_{1}T)^{-1}$\ barely
changes and then quickly reaches a constant below $T_{0}$, behaving as a
paramagnetic metal.
This completely unexpected behavior in both $^{75}\mathcal{K}$ and
$(T_{1}T)^{-1}$ confirms that the transition at $T_{0}$ in Sr$_{2}$VO$_{3}%
$FeAs is unlike any transitions observed in FeSCs so far.

Let us now discuss possible orders established below $T_{0}$. First of all, we
can eliminate the usual suspects: stripe,
double-Q\cite{avci14,bohmer15a,allred16,wasser15}, and bicollinear\cite{ma09}
AFM orders, observed in other FeSCs. In the first case $\mathbf{Q}=(\pi,0)$ and
the Fe spins aligned along the $a$ axis ($s\parallel a$) generate a hyperfine
field $H_{hf}$ $\sim$ 1.5 T along the $c$ axis. This would be visible in the
$^{75}$As\ NMR spectra as a peak splitting of $\sim$ 10 MHz for $H\parallel
c$, which is far larger than the FWHM of our spectra ($\sim$ 0.05 MHz) and
easily detectable. Similarly, for $s\parallel c$, a $^{75}$As\ peak splitting
is expected for $H\parallel a$. Even for $s\parallel b$, in which case no
transferred $H_{hf}$ and thus no peak splitting are expected, considerable
line broadening due to the directional fluctuations of Fe spins should be seen
in experiments. Neither splitting nor broadening is observed in our
experiments (Fig. 2a). For the double-Q AFM
state\cite{avci14,bohmer15a,allred16,wasser15}, a combination of two spin
density waves with $\mathbf{Q}$ = ($\pi$,0) and (0,$\pi$), the magnetization
vanishes at one of the two Fe sublattices and is staggered in the other. Thus,
the $^{75}$As\ peak splitting is expected for either $H\parallel c$
($s\parallel a$) or $H\parallel a$ ($s\parallel c$), as discussed in the
Supplementary section 5, which can be ruled out by experiments. The
bicollinear AFM order\cite{ma09} can also be excluded with even more
confidence. In this case, the As environment is spin-imbalanced (three
neighboring Fe spins are aligned in one direction, and the fourth one in the
opposite), and already a plain exchange coupling would generate two
inequivalent As sites and thus a measurable splitting for any direction of
external fields. Similarly, other AFM orders with more complicated spin
structures, such as a plaquette AFM order, are excluded as discussed in the
Supplementary section 5. This conclusion is further supported by the absence
of the diverging behavior in $(T_{1}T)^{-1}$ across $T_{0}$ (Fig. 3).

\begin{figure}[ptb]
\centering
\includegraphics[width=1\linewidth]{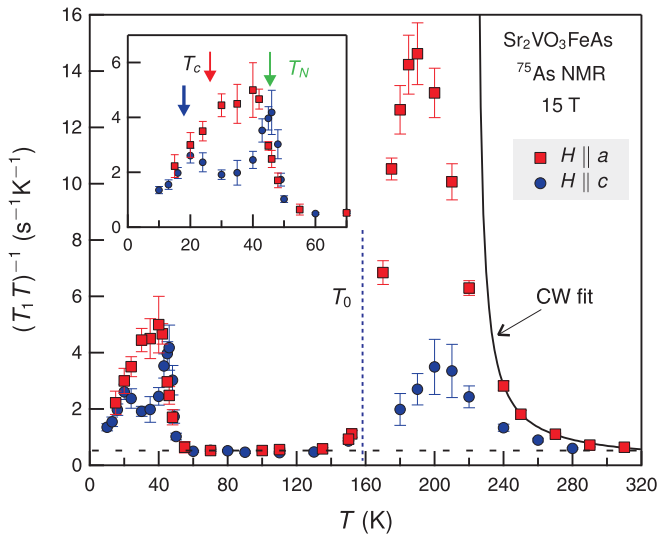}\caption{\textbf{Fe spin
fluctuations.} Temperature dependence of the $^{75}$As\ spin-lattice
relaxation rate divided by temperature $(T_{1}T)^{-1}$\ measured at 15 T. At
high temperatures, $(T_{1}T)^{-1}$\ is well described by a Curie-Weiss law
(solid line). Below $\sim240$ K, it deviates from the diverging behavior and
drops at lower temperatures, forming a large peak centered at $\sim190$ K. At
low temperatures below 120 K, $(T_{1}T)^{-1}$\ reaches a constant value
comparable to that observed at the high temperature limit, implying that the
spin fluctuations are completely gapped out. $(T_{1}T)^{-1}$\ sharply turns up
at $\sim50$ K, indicating critical slowing down of spin fluctuations
toward a magnetic order. $T_{N}\sim45$ K was identified from the sharp peak
observed for $H \parallel c$. $(T_{1}T)^{-1}$\ drops at $T_{c}$, determined by
the resistivity measurements under $H \parallel ab$ (red arrow) and $H
\parallel c$ (blue arrow), microscopically probing bulk superconductivity in
the magnetically ordered state. }
\label{fig:3}
\end{figure}

Having excluded static magnetic order, we consider now nematic or, as it is
occasionally called, vestigial partners of various AFM orders. The only
nematic order observed so far in FeSCs is the stripe-nematic order that
creates an imbalance in the orbital population between Fe $d_{xz}$ and
$d_{yz}$ states (Fig. 4b). This, in turn, induces an imbalance between As
$p_{x}$ and $p_{y}$ orbitals and dipolar in-plane anisotropy of the As Knight
shift in the twinned crystals, as observed in $e.g.$ LaFeAsO~ for $H\parallel a$ below
the nematic transition temperature\cite{fu12}.
A similar behavior is expected for the nematic partner of the bicollinear
order (Fig. 4c), which breaks the $C_{4}$ symmetry such that the (110) and
(1\={1}0) directions are not equivalent~\cite{bishop16,mazin17}. If the
generated  imbalance between the corresponding orbital Fe-$d_{xz}\pm d_{yz}$
is of the same order as in the stripe-nematic case, a peak splitting for
$H\parallel(110)$ should be detected. And, for the nematic partner of the
plaquette magnetic order, two inequivalent sites and thus a sizable splitting
are expected for every field direction. Yet, none of these signatures appears
in our $^{75}$As\ spectra for $H||a$ (100), $c$ (001) and (110) directions
(Figs. 2a, 2b and the Supplementary Fig. S4). Furthermore, our single crystal
X-ray diffraction (see Supplementary section 1), as well as the recent ARPES
study~\cite{kim15a} do not reveal any signature of a $C_{4}$ symmetry breaking.

Since the transition at $T_{0}$ retains the $C_{4}$ symmetry, and in
absence of a long range magnetic order, this transition must generate a change
in the relative occupations of the $C_{4}$ orbitals, namely $d_{xy}$,
$d_{z^{2}}$, $d_{x^{2}-y^{2}}$, and $d_{xz}\pm id_{yz}$.
Given that at high temperature we see clear indications of strong spin
fluctuations, we looked for a spin-driven scenario conserving the $C_{4}$ symmetry; a good
candidate is the vestigial (nematic) partner of the double-\textbf{Q} AFM
order\cite{fernandes16}.
It can be visualized (Fig. 4d) as a superposition of two charge/orbital
density waves with \textbf{Q}=($\pi$,0) and (0,$\pi$), which preserves the
$C_{4}$ symmetry without unit-cell doubling. This phase has a broken
translational symmetry in the Fe-only square lattice, but not in the unit cell
doubled to include the As atoms\cite{fernandes16}. Formation of the
intra-unit-cell charge/orbital density wave affects the Fe-As hybridization
and modifies the hyperfine coupling via isotropic Fermi-contact and
core-polarization interactions, accounting for the nearly isotropic
$^{75}\mathcal{K}$ Knight shift (Fig. 2). One may note that due to dipole or
orbital hyperfine interactions, the Knight shift can split for a field
parallel to (110), because half of As sites have paramagmetic neighbors along
(110), and half along (1\={1}0). However, the difference in the $d$ orbital
occupations between \textit{non}magnetic and \textit{para}magnetic Fe
are expected to be small, likely a few \% (see Supplementary section 5), in
which case the splitting will be below detection, consistent with our experiments.

If we assume nonmagnetic origin, another plausible candidate could be an
orbital-selective Mott transition. In this case, the most correlated Fe
orbital state, likely $d_{xy}$, experiences a Mott-Hubbard transition,
becoming essentially gapped, while the other orbitals remain itinerant. The
resulting occupation change in the $d_{xy}$ state of all Fe sites (Fig. 4e),
uniformly changes the hyperfine field at the As sites, retaining the $C_{4}$
symmetry and explaining the nearly isotropic change of $^{75}\mathcal{K}$
(Fig. 2). Indeed a possibility of such transition has been discussed, but,
admittedly, not in undoped pnictides, but in more strongly correlated
chalcogenides~\cite{mottkfese} and (strongly underdoped) KFe$_{2}$As$_{2}$~\cite{mottfesc1,mottfesc2}.

\begin{figure}[ptb]
\label{fig:4}\centering
\includegraphics[width=1\linewidth, bb=47 505 506 767]{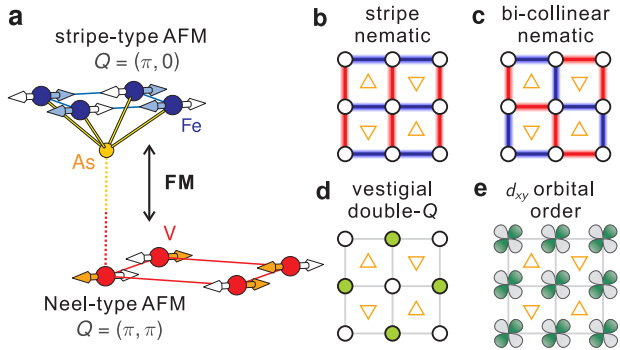}\caption{\textbf{Fe-V
interfacial spin matching and possible orders retaining a $C_{4}$ symmetry
without long-range magnetism.} \textbf{\sffamily a}, Stripe-type and Neel-type
AFM fluctuations of Fe and V spins, respectively, at high temperatures. These
different types of AFM fluctuations are frustrated via Fe-V spin
coupling developed at low temperatures. \textbf{\sffamily b-d}, various
vestigial ordered phases resulting from melting the corresponding magnetic
orders: the typical stripe nematic ($x/y$ symmetry broken), the bicollinear
nematic ($(x+y)/(x-y)$ \textit{and} the translation symmetry
broken)~\cite{mazin17}, and the vestigial double-$Q$ phase (only the
translational symmetry broken)~\cite{fernandes16}. \textbf{\sffamily e},
$d_{xy}$ orbital order driven by a possible orbital-selective Mott transition.
As opposed to \textbf{\sffamily b-d}, there is no symmetry breaking at all in
this phase compared to the high-temperature phase. In \textbf{\sffamily b}
and \textbf{\sffamily c} the symmetry breaks because some bonds are
predominantly ferromagnetic (red) and some predominantly antiferromagnetic
(blue). In \textbf{\sffamily d} green circles indicate completely
nonmagnetic Fe sites, while open circles correspond to fluctuation
paramagnetic sites. The As sites above (up-triangle) and below (down-triangle)
the Fe plane are also shown. Note that only \textbf{\sffamily d} and
\textbf{\sffamily e} are consistent with the observed $C_{4}$ symmetry, as
discussed in the main text, and are our favorite candidates for the hidden
order below $T_{0}$.}
\end{figure}

As mentioned, Sr$_{2}$VO$_{3}$FeAs experiences another transition at $T_{N}$
$\approx$ 45 K, which can be identified as a spin density wave highly distinct
from the typical stripe AFM.
Indeed, $(T_{1}T)^{-1}$\ climbs sharply below 60 K ($\ll T_{0}$) for both
$H\parallel a$ and $H\parallel c$, indicating a critical slow-down of spin
fluctuations toward a magnetic ordering at $T_{N}\sim45$ K. However,
$T_{1,a}^{-1}/T_{1,c}^{-1}$ remains isotropic, suggesting that the coupling
between Fe spins and As is due to hybridization, which can only generate a
magnetic moment on As if As environment is spin-imbalanced. This excludes such
AFM orders as stripe, Neel or double-\textbf{Q}, but would be consistent with
a longer period AFM order. Also the progressive broadening of
$^{75}$As\ spectrum at low temperatures, as shown in Figs. 2a and 2c, suggests
a long wavelength, and possibly incommensurate, spin density wave. Neutron
diffraction~\cite{tegel10,hummel13}, which observed magnetic Bragg peaks at
$\mathbf{Q}=(1/8,1/8,0)$ below $T_{N}\sim45$ K, is consistent with this
conclusion, although it was incorrectly attributed to an ordering of V spins
in the previous
studies~\cite{nakamura10,cao10,tatematsu10,sefat11,munevar11,tegel10,hummel13}.
Upon further temperature lowering, $(T_{1}T)^{-1}$\ abruptly drops at $T_{c}$.
This proves that the superconducting gap opens up on the magnetic Fe sites,
and emerges on the background of the remaining, but still strong,  spin
fluctuations with a $C_{4}$ symmetry below $T_{N}$.
How spin density wave competes or cooperate with superconductivity remains as an important question.

We shall now address an essential question: what suppresses the expected
stripe order in the FeAs layer and the Neel order in the SrVO$_{3}$ layer? The
former can be suppressed via the mechanism in which Neel-type spin
fluctuations of the localized magnetic moments are coupled to the itinerant
electrons' stripe spin fluctuation~\cite{wang14}. The stripe order, with
$\mathbf{Q}$=($\pi,0$) or (0,$\pi$), is relatively fragile and can give way to
bicollinear, double-$\mathbf{Q}$, and, possibly, plaquette orders, due to AFM
fluctuation with additional $\mathbf{Q}$'s~\cite{wang14,fernandes16,glasbrenner15}. Such magnetic frustration is due
to the long range magnetic interactions, reflecting the itinerancy of Fe
electrons. Fluctuation at $\mathbf{Q}$=($\pi,\pi$), normally weak in FeSCs, can be
enhanced through coupling to the $\mathbf{Q}$=($\pi,\pi$) fluctuations of V
spins\cite{wang14} (Fig. 4a). This destabilizes the $C_{2}$ stripe AFM or
nematic orders, but encourages the $C_{4}$ symmetric vestigial charge/orbital
density wave phases~\cite{wang14,fernandes16}. Note that in Sr$_{2}%
$(Mg,Ti)O$_{3}$FeAs and Ca$_{2}$AlO$_{3}$FeAs, isostructural compounds with
nonmagnetic oxide layers the standard stripe ordering is not
suppressed~\cite{yamamoto12,kinouchi11}. Clearly, frustration of stripe Fe and Neel V spin fluctuations, via magnetic proximity coupling, is essential for inducing an unusual hidden phase in Sr$_2$VO$_3$FeAs.

The coupling between the itinerant Fe electrons and the localized V spins also
suppress the Neel order in the SrVO$_{3}$ layer. In the SrVO$_{3}$ layers, the
nearest neighbor superexchange interaction would dominate and generate a
stable Neel order. In fact, compared to other V perovskite oxides, such as
LaVO$_{3},$ SrVO$_{3}$FeAs should have $stronger$ exchange coupling, because
of the more straight V-O-V bonds. However, the measured Curie-Weiss temperature of
$T_{CW}$$\sim-100$ K in Sr$_{2}$VO$_{3}$FeAs is considerably lower than
$T_{CW}$$\sim-700$ K in LaVO$_{3}$~\cite{lvo}. The unexpectedly low $T_{CW}$
comes from an additional ferromagnetic coupling between the V spins via indirect
double-exchange-like interaction mediated by the Fe
electrons\cite{glasbrenner14}. This frustrates and weakens the V AFM
superexchange interaction suppressing the long-range V spin order at low temperatures. Indeed, in our detailed LDA+U calculations we
found that the calculated magnetic interaction is extremely sensitive to the
on-site Coulomb energy $U$ and the Hund's coupling $J$. At $U-J=5$ eV, the
superexchange interaction, which is inversely proportional to $U$, is
significantly suppressed, while the Fe-mediated one is enhanced, so that the
net magnetic interaction becomes weakly ferromagnetic in the planes. For
$U-J=4$, it changes sign and becomes antiferromagnetic, consistent with a
previous report~\cite{nakamura10}. This demonstrates that the SrVO$_{3}$ lies
on the borderline of competing phases due to a delicate balance between the
superexchange and the additional indirect interactions. At the same time,
coupling between the stripe fluctuations in the Fe plane at $Q$ = ($\pi$,0)
and Neel fluctuations in the V plane $Q$ = ($\pi,\pi$) suppresses both orders
even further~\cite{wang14} and prevents V spins form ordering. The interfacial
Fe-V interaction is again crucial for the Mott-insulating SrVO$_{3}$ layers to
remain in a nearly paramagnetic ground state. Our findings therefore manifest that the physics of FeSCs can become even richer in the proximity of other correlated systems and also offers a new avenue for exploring unusual ground state in the correlated heterostructures.

\subsection*{Methods}

Single crystals of Sr$_{2}$VO$_{3}$FeAs\ were grown using self-flux techniques
as follows. The mixture of SrO, VO$_{3}$, Fe, SrAs, and FeAs powders with a
stoichiometry of Sr$_{2}$VO$_{3}$FeAs\:FeAs = 1:2 were pressed into a pellet
and sealed in an evacuated quartz tube under Ar atmosphere. The samples were
heated to 1180$^{o}$C, held at this temperature for 80 hours, cooled slowly
first to 950$^{o}$C at a rate of 2$^{o}$C/h and then furnace-cooled. The
plate-shaped single crystals were mechanically extracted from the flux. High
crystallinity and stoichiometry are confirmed by the X-ray diffraction and
energy-dispersive spectroscopy. The typical size of the single crystals is
200$\times$200$\times$10 $\mu$m$^{3}$.

Single crystal X-ray diffraction patterns were taken using an STOE single
crystal diffractometer with image plate. Single crystal X-ray diffraction
(XRD) reveals a good crystallinity in a tetragonal structure with a =
3.9155(7) {\AA } and c = 15.608(4) {\AA }, consistent with the previous
studies on polycrystalline samples. Detailed information about single crystal
XRD can be found in the Supplementary Information.

Conventional four-probe resistance of single crystals was measured in a 14 T
Physical Property Measurement System. Single crystal magnetizations were
measured in a 5 T Magnetic Property Measurement System. The size of one
crystal was too small ($\sim$ 0.15 mg) to measure the magnetization, thus 8
pieces of Sr$_{2}$VO$_{3}$FeAs\ single crystals (1.2 mg) were stacked
together. All single crystals were carefully aligned along the c-axis or the
ab-plane.

$^{51}$V\ (nuclear spin $I$=7/2) NMR and $^{75}$As\ ($I$=3/2) NMR measurements
were carried out at external magnetic fields of 14.983 T and 11.982 T,
respectively. The sample was rotated using a goniometer for the exact
alignment along the external field. The NMR spectra were acquired by a
standard spin-echo technique with a typical $\pi$/2 pulse length 2-3 $\mu$s
and the spin-lattice relaxation rate was obtained by a saturation method.

Band structure calculations were performed using two standard codes: an
all-electron linearized augmented plane wave method implemented in the WIEN2k
package~\cite{blaha01}, and a pseudopotential VASP code~\cite{kresse96}. In
both cases the gradient-corrected functional of Perdew, Burke and Ernzerhof
was used, and special care was taken to ensure proper occupancy of V orbitals
in the LDA+U calculations.


\begin{thebibliography}{99}
\bibitem{gozar08} Gozar, A. \emph{et al.} High-temperature interface superconductivity between metallic and insulating copper oxides. Nature \textbf{455}, 782-785 (2008).
\bibitem{wu13} Wu, J. \emph{et al.} Anomalous independence of interface superconductivity from carrier density. Nat. Mater. \textbf{12}, 877-881 (2013).
\bibitem{chakhaklian06} Chakhalian, J. \emph{et al}. Magnetism at the interface between ferromagnetic and superconducting oxides, Nat. Phys. \textbf{2}, 244-248 (2006).
\bibitem{sataphthy13} Satapathy, D. K. \emph{et al}. Magnetic Proximity Effect in YBa$_2$Cu$_3$O$_7$/La$_{2/3}$Ca$_{1/3}$MnO$_3$ and YBa$_2$Cu$_3$O$_7$/LaMnO$_{3+\delta}$ Superlattices, Phys. Rev. Lett. \textbf{108} 197201 (2012).
\bibitem{driza13} Driza, N. \emph{et al}. Long-range transfer of electron-phonon coupling in oxide superlattices, Nat. Mater. \textbf{11} 675 (2012).
\bibitem{hwang12} Hwang, H. Y. \emph{et al}. Emergent phenomena at oxide interfaces. Nat. Mater. \textbf{11}, 103-113 (2012).
\bibitem{chakhaklian14} Chakhalian, J., Freeland, J. W., Millis, A. J., Panagopoulos, C. and Rondinelli, J. M. Colloquium: emergent properties in plane view: strong correlations at oxide interfaces. Rev. Mod. Phys. \textbf{86}, 1189-1202 (2014).
\bibitem {tan13} Tan, S. \emph{et al}. Interface-induced superconductivity and strain-dependent spin density waves in FeSe/SrTiO$_3$ thin films, Nat. Mater. \textbf{12}, 634-640 (2013).
\bibitem {ge15} Ge, J.-F. \emph{et al}. Superconductivity above 100 K in single-layer FeSe films on doped SrTiO$_3$, Nat Mater. \textbf{14}, 285-289 (2015).
\bibitem{lee14} Lee, J. J. \emph{et al}. Interfacial mode coupling as the origin of the enhancement of $T_c$ in FeSe films on SrTiO$_3$, Nature \textbf{515}, 245-248 (2014).
\bibitem{coh16} Coh, S., Lee, D. -H., Louie, S. G., and Cohen, M. L. Proposal for a bulk material based on a monolayer FeSe on SrTiO$_3$ high-temperature superconductor, Phys. Rev. B \textbf{93}, 245138 (2016).
\bibitem{zhu09a}  Zhu, X. \emph{et al}. Transition of stoichiometric Sr$_2$VO$_3$FeAs to a superconducting state at 37.2 K, Phys. Rev. B \textbf{79}, 220512 (2009).
\bibitem{lee10b}  Lee, K.-W. and Pickett, W. E. Sr$_2$VO$_3$FeAs: A nanolayered bimetallic iron pnictide superconductor, Europhys. Lett. \textbf{89}, 57008 (2010).
\bibitem{mazin10a} Mazin, I. I. Sr$_2$VO$_3$FeAs as compared to other iron-based superconductors, Phys. Rev. B \textbf{81}, 020507 (2010).
\bibitem{qian11a} Qian, T. \emph{et al}. Quasinested Fe orbitals versus Mott-insulating V orbitals in superconducting Sr$_2$VFeAsO$_3$ as seen from angle-resolved photoemission, Phys. Rev. B \textbf{83}, 140513 (2011).
\bibitem{nakamura10} Nakamura, H. and Machida M. Magnetic ordering in blocking layer and highly anisotropic electronic structure of high-$T_c$ iron-based superconductor Sr$_2$VFeAsO$_3$: LDA+U study, Phys. Rev. B \textbf{82}, 094503 (2010).
\bibitem{kim15a} Kim, Y. K. \emph{et al}. Possible role of bonding angle and orbital mixing in iron pnictide superconductivity: Comparative electronic
structure studies of LiFeAs and Sr$_2$VO$_3$FeAs, Phys. Rev. B \textbf{92}, 041116 (2015).
\bibitem{johnston10} Johnston, D. C. The puzzle of high temperature superconductivity in layered iron pnictides and chalcogenides, Advances in Physics \textbf{59}, 803-1061 (2010).
\bibitem{sefat11} Sefat, A. S. \emph{et al}. Variation of physical properties in the nominal Sr$_4$V$_2$O$_6$Fe$_2$As$_2$, Physica C \textbf{471}, 143-149 (2011).
\bibitem{cao10} Cao, G. -H. \emph{et al}. Self-doping effect and successive magnetic transitions in superconducting Sr$_2$VFeAsO$_3$, Phys. Rev. B \textbf{82}, 104518 (2010).
\bibitem{tatematsu10} Tatematsu, S., Satomi, E., Kobayashi, Y., and  Sato, M. Magnetic ordering in V-layers of the superconducting system of Sr$_2$VFeAsO$_3$, J. Phys. Soc. Jpn. \textbf{79}, 123712 (2010).
\bibitem{tegel10}Tegel, M. \emph{et al}. Possible magnetic order and suppression of superconductivity by V doping in Sr$_2$VO$_3$FeAs, Phys. Rev. B \textbf{82}, 140507 (2010).
\bibitem{hummel13}  Hummel, F., Su, Y., Senyshyn, A., and Johrendt, D. Weak magnetism and the Mott state of vanadium in superconducting Sr$_2$VO$_3$FeAs, Phys. Rev. B \textbf{88}, 144517 (2013).
\bibitem{munevar11}  Munevar, J. \emph{et al}. Static magnetic order of Sr$_4$$A_2$O$_6$Fe$_2$As$_2$ ($A$= Sc and V) revealed by M\"{o}ssbauer and muon spin relaxation spectroscopies, Phys. Rev. B \textbf{84}, 024527 (2011).
\bibitem{ueshima14}  Ueshima, K. \emph{et al}. Magnetism and superconductivity in Sr$_2$VFeAsO$_3$ revealed by $^{75}$As- and $^{51}$V-NMR under elevated pressures, Phys. Rev. B \textbf{89}, 184506 (2014).
\bibitem{huang08}  Huang, Q. \emph{et al}. Neutron-diffraction measurements of magnetic order and a structural transition in the parent BaFe$_2$As$_2$ compound of FeAs-based high-temperature superconductors, Phys. Rev. Lett. \textbf{101}, 257003 (2008).
\bibitem{kitagawa08}  Kitagawa, K., Katayama, N., Ohgushi, K., Yoshida, M., and Takigawa, M. Commensurate itinerant antiferromagnetism in BaFe$_2$As$_2$: $^{75}$As-NMR studies on a self-flux grown single crystal, J. Phys. Soc. Jpn. \textbf{77}, 114709 (2008).
\bibitem{fu12}  Fu, M. \emph{et al}. NMR search for the spin nematic state in a LaFeAsO single crystal, Phys. Rev. Lett. \textbf{109}, 247001 (2012).
\bibitem{avci14}  Avci, S. \emph{et al}. Magnetically driven suppression of nematic order in an iron-based superconductor, Nat Commun \textbf{5}, 3845 (2014).
\bibitem{bohmer15a} B\"{o}hmer, A. E. \emph{et al}. Superconductivityinduced re-entrance of the orthorhombic distortion in Ba$_{1-x}$K$_x$Fe$_2$As$_2$, Nat. Commun. \textbf{6}, 7911 (2015).
\bibitem{allred16}  Allred, J. M. \emph{et al}. Double-Q spin-density wave in iron arsenide superconductors, Nat. Phys. \textbf{12}, 493-498 (2016).
\bibitem{wasser15} Wa$\beta$er, F. \emph{et al}. Spin reorientation in Ba$_{0.65}$Na$_{0.35}$Fe$_2$As$_2$ studied by single-crystal neutron diffraction, Phys. Rev. B \textbf{91}, 060505 (2015).
\bibitem{ma09} Ma, F., Ji, W., Hu, J., Lu, J.-Y., and Xiang, T. First-Principles calculations of the electronic structure of tetragonal $\alpha$-FeTe and $\alpha$-FeSe crystals: Evidence for a bicollinear antiferromagnetic order, Phys. Rev. Lett. \textbf{102}, 177003 (2009).
\bibitem{bishop16} Bishop, C. B., Herbrych, J., Dagotto, E., and Moreo, A. Possible bicollinear nematic state with monoclinic lattice distortions in iron telluride compounds, arXiv 1704-03495 (2017).
\bibitem{mazin17} Zhang, G., Glasbrenner, J. K., Flint, R., Mazin, I. I., and Fernandes, R. M. Double-stage nematic bond ordering above double stripe magnetism: Application to BaTi$_2$Sb$_2$O, Phys. Rev. B \textbf{95}, 174402 (2017).
\bibitem{fernandes16} Fernandes, R. M., Kivelson, S. A., and Berg, E. Vestigial chiral and charge orders from bidirectional spin-density-waves: Application to the iron-based superconductors, Phys. Rev. B \textbf{93}, 014511 (2016).
\bibitem{mottkfese} Yu, R., and Si, Q. Orbital-selective Mott phase in multiorbital models for alkaline iron selenides K$_{1-x}$Fe$_{2-y}$Se$_2$, Phys. Rev. Lett. \textbf{110}, 146402 (2013).
\bibitem{mottfesc1} Medici, L., Giovannetti, G., and Capone, M. Selective Mott physics as a key to iron superconductors, Phys. Rev. Lett. \textbf{112}, 177001 (2014).
\bibitem{mottfesc2} Misawa, T., Nakamura, K., and Imada, M. Ab initio evidence for strong correlation associated with Mott proximity in iron-based superconductors, Phys. Rev. Lett. \textbf{108}, 177007 (2012).
\bibitem{wang14} Wang, X. and Fernandes, R. M. Impact of local-moment fluctuations on the magnetic degeneracy of iron-based superconductors, Phys. Rev. B \textbf{89}, 144502 (2014).
\bibitem{glasbrenner15} Glasbrenner, J. K. \emph{et al}. Effect of magnetic frustration on nematicity and superconductivity in iron chalcogenides, Nat. Phys. \textbf{11}, 953 (2015).
\bibitem{yamamoto12} Yamamoto, K. \emph{et al}. Antiferromagnetic order and superconductivity in Sr$_4$(Mg$_{0.5-x}$Ti$_{0.5+x}$)$_2$O$_6$Fe$_2$As$_2$ with electron doping: $^{75}$As-NMR Study, J. Phys. Soc. Jpn. \textbf{81}, 053702 (2012).
\bibitem{kinouchi11} Kinouchi, H. \emph{et al}. Antiferromagnetic spin fluctuations and unconventional nodeless superconductivity in an iron-based new superconductor (Ca$_4$Al$_2$O$_{6-y}$)(Fe$_2$As$_2$): $^{75}$As nuclear quadrupole resonance study, Phys. Rev. Lett. \textbf{107}, 047002 (2011).

\bibitem{lvo} Mahajan, A. V., Johnston, D. C., Torgeson, D. R., and Borsa, F. Magnetic properties of LaVO$_3$, Phys. Rev. B \textbf{46}, 10966 (1992).
\bibitem{glasbrenner14} Glasbrenner, J. K., \v{Z}uti\'{c}, I., and Mazin, I. I. Theory of Mn-doped II-II-V semiconductors, Phys. Rev. B \textbf{90}, 140403 (2014).
\bibitem{blaha01} Blaha, P., Schwarz, K., Madsen, G. K. H., Kvasnicka, D. and Luitz, J. \textit{Wien2K: An augmented plane wave + local orbitals program for calculating crystal properties} (Techn. Universit\"{a}t Wien, Wien, 2001).
\bibitem{kresse96} Kresse G., and Furthm\"{u}ller, J. Efficient iterative schemes for ab initio total-energy calculations using a plane-wave basis set, Phys. Rev. B \textbf{54}, 11169 (1996).


\end{thebibliography}



\let\origdescription\description
\renewenvironment{description}{
\setlength{\leftmargini}{0em}
\origdescription
\setlength{\itemindent}{0em}
\setlength{\labelsep}{\textwidth}
}
{\endlist}

\begin{description}
\item[Acknowledgment] The authors thank L. Boeri, R. Fernandes, S. Backes, R. Valenti, C. Kim, Y. K.
Kim, J.-H. Lee and K. H. Kim for fruitful discussion. This work was supported
by the NRF through the Mid-Career Researcher Program (No. 2012-013838), SRC
(No. 2011-0030785), the Max Planck-POSTECH Center for Complex Phase Materials
(No.2011-0031558) and also by IBS (No.IBS-R014-D1-2014-a02). The work at IFW
Dresden has been supported by by the Deutsche Forschungsgemeinschaft (Germany)
via DFG Research Grants BA 4927/1-3 and the Priority Program SPP 1458.
Financial support through the DFG Research Training Group GRK 1621 is
gratefully acknowledged. J.H.P was also supported by the National Creative
Initiative (No. 2009-0081576). I.M acknowledges funding from the Office of
Naval Research (ONR) through the Naval Research Laboratory's Basic Research
Program, and from the A. von Humboldt Foundation.

\item[Author Contributions] J.S.K, J.M.O, and S.H.B conceived the experiments.
J.M.O synthesized the samples. J.M.O carried out the transport and
magnetization measurements. S.H.B and B.B contribute to the NMR measurements
and the analysis. C.H, R.K.K. S.Y.P, S.D.J and J.H.P contribute to single
crystal X-ray diffraction measurements. I.M, S.I.H, J.H.S, Y.B and E.G.M contribute
to the theoretical calculations and the analysis. J.M.O, S.H.B, E.G.M, I.M and
J.S.K co-wrote the manuscript. All authors discussed the results and commented
on the paper.

\item[Competing financial interests] The authors declare no competing
financial interests.

\item[Additional information] Correspondence and requests for materials should
be addressed to S.-H. Baek~(email: sbaek.fu@gmail.com) and J. S. Kim~(email: js.kim@postech.ac.kr).
\end{description}

\pagebreak

\end{document}